\documentclass[aps,prd,reprint,noshowpacs,twocolumn,amsmath,amssymb]{revtex4}
\usepackage{epsfig}
\usepackage{ifpdf}
\usepackage{graphicx}
\usepackage{amssymb,amsmath,lineno,mathbbol,upgreek}
\usepackage[tight]{subfigure}
\usepackage{mathrsfs}
\usepackage{bbm}
\usepackage{slashed}
\usepackage{natbib}

\usepackage{calligra}
\DeclareMathAlphabet{\mathcalligra}{T1}{calligra}{m}{n}
\DeclareFontShape{T1}{calligra}{m}{n}{<->s*[2.2]callig15}{}

\usepackage{float}
\usepackage{afterpage}
\usepackage{float}
\usepackage{lipsum}

\usepackage{calligra}
\DeclareMathAlphabet{\mathcalligra}{T1}{calligra}{m}{n}
\DeclareFontShape{T1}{calligra}{m}{n}{<->s*[2.2]callig15}{}


\begin{document}

\bibliographystyle{prsty}

\title{Quantum Berezinskii-Kosterlitz-Thouless transition in the superconducting phase of (2+1)-dimensional quantum chromodynamics}

\author{Laith H. Haddad}
\affiliation{Department of Physics, Colorado School of Mines, Golden, CO 80401,USA}
\date{\today}

\begin{abstract}
We study superconductivity in the hadron-quark mixed phase of planar quantum chromodynamics (QCD) within the large $N$ limit of a Gross-Neveu model modified by a repulsive vector term. At high densities, the combination of scalar attraction and repulsive space-like part of the vector interaction squeezes quarks into baryonic composite states, i.e., Dirac fermions with even numbers of bosonic vortices attached. The time-like vector component induces Cooper pairing between these Fermi surface modes. Remarkably, at zero temperature, competition between the quark density and mass destroys superconductivity via a Berezinskii-Kosterlitz-Thouless (BKT) phase transition driven by diverging chiral quantum fluctuations near criticality. Dissolution of logarithmically bound singlet diquarks is catalyzed by in-plane chiral mixing associated with $\mathbb{Z}_2 \otimes \mathbb{Z}_2 \to \mathbb{Z}_2$ chiral symmetry breaking of the Fermi surface into a transverse spin-polarized triplet ground state. We calculate the QCD phase diagram for quark chemical potential above the baryon mass based purely on Fermi surface considerations and find good agreement with results obtained by other methods. We address similarities between our quantum BKT transition and those found using holographic techniques. 
\end{abstract}

\maketitle

\section{Introduction}
  
Presently, a thorough knowledge of the phases of quantum chromodynamics (QCD) is lacking and constitutes a significant hurdle towards our understanding of the rich phenomenology encoded within the standard model of particle physics~\cite{Alford2008,Fukushima2011,Weise2012}. Although QCD is notoriously difficult to solve directly, much insight has been gained through a combination of laboratory experimentation, observational cosmology, and effective theoretical models. For instance, data from heavy ion collision experiments tells us that for temperatures above the QCD scale, i.e., $\Lambda_\mathrm{QCD} \sim 200 \,  \mathrm{MeV}$, and at low quark densities the physics is described by a quark-gluon plasma. In contrast, we know considerably less about the physics at low temperatures and increasingly higher densities beginning near the baryon density, $\Lambda_\mathrm{QCD}^3 \sim 1 \, \mathrm{fm}^{-1}$. Although, it is true that at extremely high densities perturbative asymptotic freedom applies, as hadrons ultimately dissolve into degenerate quark matter, asymptotic freedom does not apply for the intermediate regimes. Adding to the problem, first principle lattice simulations are impractical here due to the infamous sign problem and empirical data is constrained by the fact that the required quark densities are unreachable in the laboratory, occurring only in the deep interior of dense stelar objects such as neutron stars. Nevertheless, various phenomenological methods such as the Nambu-Jona-Lasinio (NJL)~\cite{1nambu61,2nambu61} and random matrix models~\cite{Stephanov1996,Halasz1997}, among others, have lead to conjectured exotic phases including color superconductivity (CSC), color-flavor locking, and mixed phases characterized by the simultaneous coexistence of hadrons and quarks~\cite{Fukushima2011}.

In this article, we investigate the interplay between chiral symmetry breaking (CSB) and color superconductivity within the hadron-quark mixed phase of (2+1)-dimensional QCD. We model the intermediate-to-large quark density regime of the QCD phase diagram within the large $N$ semiclassical limit of the Gross-Neveu-Thirring model~(see for example~\cite{Andrianov1992} and references therein). This model couples $N$ species of fermions through scalar (Gross-Neveu) and vector (Thirring) interactions. As we are concerned primarily with intermediate quark densities, a desirable feature of our model is the presence of terms that are equivalent to semiclassical corrections coming from an external gauge field. This accounts for the early onset of asymptotic freedom associated with perturbative QCD that applies towards higher densities.

The character of the interactions in our model are as follows. The scalar term is taken to be attractive in order to allow for chiral condensation $\langle \bar{q} q \rangle \ne 0$. In contrast, we have chosen a repulsive vector term. This is justified from multiple lines of reasoning~\cite{Kitazawa2002,Buballa2005,Abuki2009,Coelho2010,Weissenborn2012,Masuda2013,Orsaria2014,Menezes2014,Ferrer2015}. For example, both the instanton-anti-instanton molecule model and the renormalization group approach result in repulsive vector corrections in their respective effective actions. From an empirical standpoint, the equation of state for neutron stars requires a degree of stiffness that seems to be accounted for only by adding a repulsive vector term to standard NJL-type theories. The presence of a repulsive vector interaction plays a significant role in the formation of a color superconducting phase. In particular, a space-like metric signature produces a negative vector density interaction $-(\bar{\psi}\gamma^0\psi )^2$, which gives an attractive quark-quark channel and hence the possibility of diquark condensation $\langle q q \rangle \ne 0$ associated with CSC.

We will demonstrate how the interplay between attraction and repulsion in our model leads to formation of the Fermi surface at low temperatures by way of a Landau-Ginzburg-Wilson mechanism involving the amplitude of the quark field. In essence, this is due to the repulsive space-like part of the vector interaction $(\bar{\psi}\gamma^i\psi )^2$. It is important to emphasize that in the full picture this occurs independent of, yet simultaneous with, the formation of bilinear condensates and hence chiral symmetry breaking. Pauli blocking combined with the ensuing finite expectation value for the quark amplitude forces the internal spinor phase to cycle around the circle in order to maintain a zero overall expectation for the quark field. Thus, we find that excitations of the Fermi surface are composite quarks: fermions with even numbers of bosonic vortices attached. The extra bosonic degrees of freedom effectively mitigate the repulsive exchange interaction between quarks. The net result is a large quark density stiffness consistent with what we know about the properties of dense QCD from neutron star considerations~\cite{Alford2009}.

 Our model exhibits a breakdown in the superconducting phase through an infinite-order Berezinskii-Kosterlitz-Thouless (BKT) type phase transition. Significantly, this occurs as a conventional finite temperature BKT transition along a boundary that intersects the chemical potential axis at zero temperature, hence surviving as a \emph{quantum} BKT transition (QBKT). Detailed analysis reveals the QBKT mechanism to be fundamentally rooted in competition between the scalar part of the interaction and the time-like vector density. The former generates a quark mass which mimics an external magnetic field perpendicular to the plane by introducing spin polarization asymmetry, while the latter generates a finite quark density. The vortex degrees of freedom in our BKT transition are precisely the composite quarks that result from squeezing the quark amplitude and constitute the Cooper pairs near the Fermi surface. As we reduce the quark chemical potential (equivalently quark density) through the BKT transition the system favors quark-anti-quark over diquark condensation. Consequently, we find that diquark dissociation at BKT criticality occurs precisely when Cooper pairs transform from singlet to triplet states.

 The quantum BKT transition in $(2+1)$d previously found for a D3/D5 brane system~\cite{Jensen2010} using holography~\cite{Witten1998,Maldacena1999} is of particular relevance to the results presented in the present article. The initial holographic results were subsequently extended to include various other driving parameters and systems~\cite{Jensen2010.2,Iqbal2010,Evans2011.1,Evans2011.3,Grignani2012,Bigazzi2012}. The QBKT transition in~\cite{Jensen2010} was found to be driven by competition between applied magnetic field strength and quark density, in contrast to standard thermally driven BKT. This scenario bears a remarkable resemblance to our problem wherein one finds QBKT to be driven by competition between the quark mass and chemical potential, where the effect of the former on the quark spin polarization is analogous to that of applied magnetic field and the latter directly proportional to the quark density. This parallel attests to the richness of four-fermion theories. Indeed, in this article we show that the four-fermion approach has the advantage of providing deeper insight by elucidating the microphysics underlying the BKT mechanism, in opposition to holography which defines the boundary field theory purely in terms of its bulk gravity dual at the unfortunate expense of a clear physical description for the BKT mechanism.

This article is organized as follows. Section~\ref{Model} establishes the foundation of our model by specifying the Lagrangian density for attractive and repulsive four-fermion interactions. In Sec.~\ref{FermiSurface}, we derive the structure of the Fermi surface based on low-temperature symmetry breaking, treating the quark mass as an external tunable parameter. In Sec.~\ref{QFluctuations}, we interpret excitations of the Fermi surface in terms of composite quarks. Here we uncover the properties of quasi-quarks which result directly from the reduced compressibility of the system induced by quark-quark repulsion. In Sec.~\ref{QBKT} we show that the dissociation of Cooper pairs in the superfluid state proceeds via the Berezinskii-Kosterlitz-Thouless mechanism at both finite as well as zero temperature. In Sec.~\ref{PhaseDiagram} we connect to well-known established results by constructing, based on our model, the temperature-chemical potential phase diagram for QCD. In Sec.~\ref{Conclusion}, we conclude.

\section{The model}
\label{Model}

We begin our analysis working directly from the microscopic dynamics of quark degrees of freedom interacting locally through Lorentz scalar-scalar and vector-vector terms. We compute the properties of the CSC diquark condensate from the resulting bound states of quarks of the same chirality. The Lagrangian density for our model is given by
\begin{eqnarray}
\mathcal{L} = \mathcal{L}_0 +  \mathcal{L}_S +  \mathcal{L}_V \, , \label{FullLagrangian}
\end{eqnarray}
with the kinetic, scalar, and vector contributions given explicitly by
\begin{eqnarray}
\mathcal{L}_0 &=&  \sum_{n=1}^N \bar{\psi}^{(n)} \left( i \gamma^\mu  D_\mu - m  + \mu \gamma^0  \right)   \psi^{(n)} \, ,  \label{kinetic} \\
 \mathcal{L}_S &=&  \frac{g_S^2 }{2}   \left(  \sum_{n=1}^N \bar{\psi}^{(n)}   \psi^{(n)} \right)^2 \, ,  \label{scalar} \\ 
 \mathcal{L}_V &=&  - \frac{ g_V^2}{2}  \left(  \sum_{n=1}^N \bar{\psi}^{(n)}   \gamma^\mu \psi^{(n)}  \right)^2  \, , \label{vector}
\end{eqnarray}
which includes explicit quark mass $m$ and chemical potential $\mu$. The superscript index refers to $N$ species of quarks with scalar and vector couplings scaling like $g_S^2 , \, g_V^2\sim 1/N$. This $N$-dependent scaling of the couplings will be important when we consider the semiclassical regime. It follows that in the special case $\mu = m = 0$, Eq.~(\ref{FullLagrangian}) is symmetric under an $\mathrm{SU}(N)_R$ $\otimes$ $\mathrm{SU}(N)_L$ chiral transformation. In the following we take the couplings to satisfy $g_S^2 = 3 g_V^2 \equiv g^2$, as there is no a priori relationship between these and this form leads to a spin-symmetric total interaction. Moreover, we will suppress the summation over $n$, neglecting the species index, in which case Eq.~(\ref{FullLagrangian}) has an explicit discrete $\mathbb{Z}_4$ chiral symmetry. The covariant derivative $D_\mu = \partial_\mu -  i e A_\mu$ in Eq.~(\ref{kinetic}) couples the quark field to a non-dynamical gauge field, as no associated field strength $\mathrm{tr}(F_{\mu \nu})^2$ appears in the action. The reason for including this gauge potential will be made clear shortly. We work throughout in the two-dimensional Weyl formalism $\psi = (\psi_1, \, \psi_2 )^T$ using the first two Pauli matrices for the space-like part of our theory and the third one for the time-like part.

 An interesting feature of the model Eqs.~(\ref{FullLagrangian})-(\ref{vector}) is that the large $N$ limit of the time-like part of the vector interaction mimics weak coupling to a semiclassical gauge potential. To see this, consider a weak gauge interaction treated perturbatively and semi-classically by retaining only the scalar part of the four-vector potential. Expanded in a zeroth-order local static contribution plus a first-order nonlocal part, this four-vector potential reads 
\begin{eqnarray}
A_\mu(| {\bf r} - {\bf r}'|) \approx  \left[ A_0(| {\bf r} - {\bf r}' |), \, 0 \, , 0 \, \right]^T \, , 
\end{eqnarray}
with 
\begin{eqnarray}
&&\hspace{-1pc}A_0(| {\bf r} - {\bf r}' |) \approx  \\
&&\hspace{2pc}(1/\alpha)  A_0^{(0)} \,  \delta(| {\bf r} - {\bf r}'|) + (1/\alpha^2)  A_0^{(1)} \, f(| {\bf r} - {\bf r}'|) \, , \nonumber 
\end{eqnarray}
where $\alpha$ is the fine structure constant. The first term produces a point-like (mean-field) correction to the chemical potential and the second term a nonlocal correction. An effective chemical potential $\tilde{\mu}$ can then be defined as
\begin{eqnarray}
&&\hspace{-1pc}\tilde{\mu}(| {\bf r} - {\bf r}' |) \equiv \\
&&\hspace{0pc} \left[  \mu +  (1/\alpha)  A_0^{(0)} \right]  \,  \delta(| {\bf r} - {\bf r}'|) + (1/\alpha)^2 A_0^{(1)} \, f(| {\bf r} - {\bf r}'|)\, . \nonumber 
\end{eqnarray}
But, the zeroth-order correction is identical to a Hartree-Fock term generated by condensation in the time-like vector density $\langle \psi^\dagger \psi \rangle$. Hence, coupling to a gauge field is actually generated internally by the vector interaction and remains non-dynamical since the required field strength does not appear in $\mathcal{L}$. Thus, promoting the kinetic term in Eq.~(\ref{kinetic}) to the status of covariance is warranted.  

 The Lagrangian Eq.~(\ref{FullLagrangian}) possesses a convenient symmetry, advantageous for modeling the hadron-quark mixed phase. First, consider that one may view the parameter $m$ as a dynamical scalar mass generated by a finite chiral condensate $m = g^2  \langle \bar{\psi} \psi \rangle + m_0$, where $m_0$ is the bare mass. In this picture the chiral condensate $\langle \bar{\psi} \psi \rangle$ is a Hartree-Fock mean field term generated by the attractive scalar part of Eq.~(\ref{FullLagrangian}). In a similar fashion, as explained above, one may view the gauge potential $A_0$ as a diquark condensate generated by the attractive time-like vector density $\langle \psi^\dagger  \psi \rangle$ through $A_0 = - (g^2/3)  \langle \psi^\dagger \psi \rangle + A_0^{(0)}$, with $A_0^{(0)}$ the bare potential associated with the diquark bound state. Focusing on the density dependent parts, transforming between the quark-antiquark and diquark representations of $\mathcal{L}_\mathrm{int}$, the terms $- m$ and $(\mu + A_0) \gamma^0$ transform like 
\begin{eqnarray}
\mathrm{(quark\!-\!antiquark)} && \hspace{-2pc}   \hspace{4pc}    \mathrm{(diquark)} \nonumber  \\
\hspace{-3.75pc} -m:  \hspace{1.15pc} -  \langle \bar{\psi} \psi \rangle  \hspace{2pc} &\longrightarrow&  \hspace{1pc}   -  \langle \psi^\dagger \gamma^0 \psi \rangle  \gamma^0  \, , \\
\hspace{-1.25pc} ( \mu + A_0) \gamma^0:  \hspace{.5pc}  -  \langle \bar{\psi} \gamma^0 \psi \rangle \gamma^0   \hspace{1pc}   &\longrightarrow&  \hspace{2pc}  -  \langle \psi^\dagger \psi \rangle  \, . 
\end{eqnarray}
Thus, the roles of the mass and chemical plus gauge potential are switched but their combination is invariant under transformations between the meson and quark matter representations. We will see that it is the competition between these two terms that is fundamentally responsible for dissolution of the diquark state. All of this strengthens the case for investigating Eqs.~(\ref{FullLagrangian})-(\ref{vector}) as a plausible model for the coexistence phase of QCD near the superconducting transition.

\section{Chiral Symmetry Breaking: Formation of the Fermi surface}
\label{FermiSurface}

The formation of a sharp Fermi surface in our model is closely related to competition between short range quark repulsive and attractive terms in Eq.~(\ref{FullLagrangian}). In this section we track the development of the Fermi surface at low temperatures by treating the quark mass as an external tunable parameter, revealing in the process chiral symmetry breaking of the discrete $\mathbb{Z}_4$ symmetry of Eq.~(\ref{FullLagrangian}). It is important to emphasize that in our treatment, chiral symmetry breaking is essentially generated within a Landau-Ginzburg-Wilson type framework for the quark field. We will see that at high densities the resulting finite expectation for the quark amplitude forces stringent constraints on the internal phase of the quark spinor due to Pauli blocking for Fermi surface states.

Formation of the Fermi surface is demonstrated by first expanding the spinor field in $\mathcal{L}$ in terms of single-particle Dirac states as
\begin{eqnarray}
\psi({\bf r}, t) =  \sum_{i} \left[   c_{1, i}(t)  \chi_{1, i}({\bf r})  , \,  c_{2, i}(t)  \chi_{2, i}({\bf r})    \right]^T\, , \label{decomp1}
\end{eqnarray}
and
\begin{eqnarray}
\psi^\dagger({\bf r}, t) = \sum_{i} \left[   c_{1, i}(t)  \chi_{1, i}^*({\bf r})  , \,  c_{2, i}(t)  \chi_{2, i}^*({\bf r})    \right] \, , \label{decomp2}
\end{eqnarray}
where the summation index labels the single-particle states. We let the spinor functions $\psi_i({\bf r})$ $=$ $[ \chi_{1, i}({\bf r})  ,$$ \,  \chi_{2, i}({\bf r})   ]^T$ solve the time-independent massless Dirac equation 
\begin{eqnarray}
\left(  i     {\bf \sigma} \cdot    \nabla  - \epsilon_i  \right)   \psi_i({\bf r}) = 0 \, , 
\end{eqnarray}
with eigenvalues $\epsilon_i$. Using orthonormality $\int \!  d{\bf r} \,  \bar{\psi}_i({\bf r})  \psi_j({\bf r})$ $=$ $\delta_{i j}$, substituting the field expansions into the action, and performing the spatial integral, we obtain the action $S = S_0 + S_\mathrm{int}$ with 
\begin{eqnarray}
         \hspace{-1pc}&&S_0 =  \int_0^{\hbar \beta} \!  dt   \sum_{i}  \left\{   c_{1,i}^*(t)  \partial_t c_{1,i}(t) +    c_{2,i}^*(t) \partial_t c_{2,i}(t)  \right.\nonumber  \\
         \hspace{-1pc}&& \hspace{6pc}+ \left.  \epsilon_i   |c_{1,i}(t)|^2 +   \epsilon_i   |c_{2,i}(t)|^2 \right. \nonumber  \\
       \hspace{-1pc}&& \hspace{2pc}+ \left.  \left[ m    - \tilde{\mu}(T)  \right]   | c_{1,i}(t)|^2 - \left[  m    +  \tilde{\mu}(T) \right]  |c_{2,i}(t)|^2  \right\} \, ,   \nonumber \\
         \hspace{-1pc}&& S_\mathrm{int} =  \int_0^{\hbar \beta} \!  dt  \sum_{i, j, k. l}  \frac{\bar{g}^2_{i j k l }}{2}  \left[   c_{1,i}^*(t) c_{1,j}^*(t)   c_{1,k}(t)  c_{1,l}(t)\right. \nonumber \\
                      \hspace{-1pc}&& \hspace{6pc}+ \left.   c_{2,i}^*(t) c_{2,j}^*(t)   c_{2,k}(t)  c_{2,l}(t) \right]    \, ,   \label{brokensymmetry1}
\end{eqnarray}
where the matrix for the spatially renormalized interaction strength is 
\begin{eqnarray}
\bar{g}^2_{i j k l } \equiv g^2 \int \! d{\bf r} \,  \chi_{1, i}^*({\bf r}) \chi_{1, j}^*({\bf r})  \chi_{1, k}({\bf r}) \chi_{1, l}({\bf r})\, , 
\end{eqnarray}
with a similar definition for the lower spin component. Note that the temperature appears in the integration limits through $\beta = (k_B T)^{-1}$. With exponential forms for the coefficients $c_{1(2),i}(t)  = c_{1(2),i} \, \mathrm{exp}( - i \epsilon_i t /\hbar)$, at critical temperatures $T_{c, i}$ the combined quadratic and quartic terms develop local minima away from $c_{1,i}, c_{2,i} = 0$, whose distances from the origin in spin space are eignevalue dependent. A significant point here is that, since we are interested in a finite quark density at low temperatures, the active modes are those lying nearest to the Fermi surface with energy $\sim \epsilon_F$. The expansions in Eq.~(\ref{brokensymmetry1}) can then be approximated by sums over the single index $n_{\epsilon_F}$.

We should not be alarmed by the development of a finite expectation for the coefficients $c_{1(2)}$, as these define the overall and relative spinor amplitudes $\rho^{1/2} =  \sqrt{ c_1^2 |\chi_1|^2 +  c_2^2 |\chi_2|^2}$ and $\rho_\mathrm{rel}^{1/2} =  \sqrt{| c_1^2 |\chi_1|^2 -  c_2^2 |\chi_2|^2|}$, respectively, the latter encoding the spin polarization. Since $c_{1(2)}$ are real, the phase structure of the fermion field $\psi$ resides in the space and time dependence of the decomposition Eqs.~(\ref{decomp1})-(\ref{decomp2}), so that we correctly obtain $\langle \psi \rangle = 0$ even though $\langle c_{1(2)} \rangle \ne 0$. We will this point in detail in Sec.~\ref{QFluctuations}. At the Fermi surface the Hessian matrix for Eq.~(\ref{brokensymmetry1}) can then be computed which yields four regimes defined by the following inequalities: 

\begin{enumerate}

\item  for $m    + \tilde{\mu}  - \epsilon_F$$<$$0$ and $- m + \tilde{\mu}  - \epsilon_F$$>$$0$, two minima occur at $c_2$$=$$ \pm \sqrt{ (  \epsilon_F - \tilde{\mu} +  m    )/g^2}$, $c_1$$=$$0$;

\item  for $m    + \tilde{\mu}  - \epsilon_F$$> $$0$ and $- m     + \tilde{\mu}  - \epsilon_F$$<$$0$, two minima occur at $c_2$$=$$0$, $c_1$$=$$\pm \sqrt{ (  (  \epsilon_F - \tilde{\mu} +  m    )/g^2}$;

\item for $m    + \tilde{\mu}  - \epsilon_F$$< $$0$ and $- m     + \tilde{\mu}  - \epsilon_F < 0$, four minima occur at $c_2$$=$$\pm \sqrt{ ( \tilde{\mu} - m    - \epsilon_F)/g^2}$, $c_1$$=$$\pm \sqrt{ ( \tilde{\mu} + m    - \epsilon_F)/g^2}$;

\item for $m    + \tilde{\mu}  - \epsilon_F$$>$$0$ and $- m     + \tilde{\mu}  - \epsilon_F$$>$$0$, four minima occur at $c_2$$=$$\pm \sqrt{ ( \tilde{\mu} + m    - \epsilon_F)/g^2}$, $c_1$$=$$\pm \sqrt{ ( \tilde{\mu} - m    - \epsilon_F)/g^2}$.

\end{enumerate}

\begin{figure}[]
\centering
\subfigure{
\includegraphics[width=.54\textwidth]{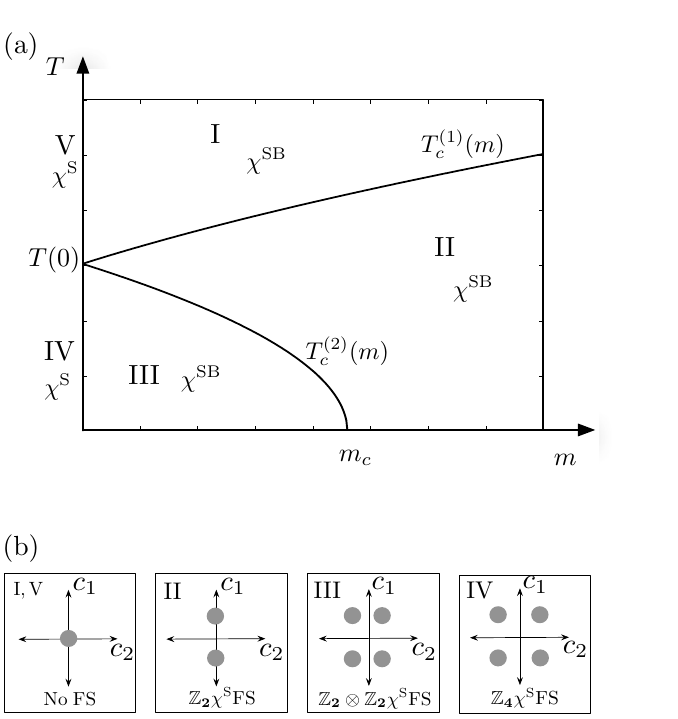}} \\
\caption[]{\emph{Chiral symmetry breaking associated with formation of the Fermi surface}. (a) The temperature-mass plane phases in the development of the Fermi surface are separated by Landau-Ginzburg-Wilson type transitions. Chiral symmetry of the full theory in each region is indicated as either broken ($\chi^\mathrm{SB}$) or retained ($\chi^\mathrm{S}$). In regions I and V ($T > T_c^{(1)}$), no well-defined Fermi surface exists. In region II ($T_c^{(2)}< T< T_c^{(1)}$), the Fermi surface is formed in the presence of a large quark mass or a small mass at moderately low temperatures. Region III ($T< T_c^{(2)}$, $m \ne 0$) corresponds to a well-defined Fermi surface in the presence of a small but non-zero quark mass. In region IV ($T< T_c^{(2)}$, $m = 0$), the Fermi sea is comprised of massless quarks. (b) Minima (ground states) of the effective potential in Eq.~(\ref{brokensymmetry1}) are depicted as disks in the spin-space coordinate plane for each region in (a) with spin up and down along the vertical and horizontal axes, respectively. The inherited chiral symmetry (degeneracy) group of the Fermi surface (FS) is also indicated.   
}
\label{PseudospinDomain}
\end{figure}

To delineate the temperature dependent regimes determined by the above inequalities, we expand the chemical potential at low temperatures keeping only the first two terms. For a Fermi system $\mu(T) \approx  \mu_0  - \mu_2  T^2$, where $\mu_0$ is the chemical potential at $T$$=$$0$ and the coefficient $\mu_2$ can be determined by standard methods. The broken symmetry phase corresponding to four minima in Eq.~(\ref{brokensymmetry1}) is delineated by the condition $T < T_c^{(2)}(m)$, where the mass-dependent critical temperature is 
\begin{eqnarray}
T_c^{(2)}(m) \equiv \sqrt{ (1/\mu_2 ) \left[  \tilde{\mu}_0 - ( m + \epsilon_F ) \right]} \, , \label{CTemp2}
\end{eqnarray}
and we have defined $\tilde{\mu}_0 = \mu_0 - A_0^0$. In contrast, the phase with two minima is determined by the condition $T_c^{(2)}(m) < T < T_c^{(1)}(m)$, where 
\begin{eqnarray}
T_c^{(1)}(m) \equiv \sqrt{(1/\mu_2) \left[  \tilde{\mu}_0 +  ( m  -  \epsilon_F ) \right]} \, . \label{CTemp1}
\end{eqnarray}
Figure~\ref{PseudospinDomain} depicts the various temperature-dependent regimes involved in forming a sharp Fermi surface. Note in particular the quantum critical point $m_c = | \epsilon_F - \tilde{\mu}_0|$.

A key point in our analysis is that the critical temperatures, Eqs.~(\ref{CTemp2})-(\ref{CTemp1}), depend on the Fermi energy $\epsilon_F$. This is because we have only considered modes for the expansions in Eq.~(\ref{brokensymmetry1}) with energies near $\epsilon_F$. Higher temperatures blur the Fermi surface allowing for active modes deep within the Fermi sea. When this occurs, the structure underlying Eqs.~(\ref{CTemp2})-(\ref{CTemp1}) is smoothed out in that we can no longer identify a single effective potential for the action Eq.~(\ref{brokensymmetry1}). Instead, what we see at higher temperatures is a quantum blurred band of ``critical'' temperatures with no particular energy (or temperature) single-out as special. A sharp transition in the Fermi sea occurs only when a particular eignevalue dominates the expansions in Eq.~(\ref{brokensymmetry1}).

\section{Excitations of the Fermi Surface}
\label{QFluctuations}

The development of finite expectation for the quark amplitude requires a completely random internal phase in order to maintain the required fermion field condition $\langle \psi \rangle$ $=$ $0$. Consider a general two-spinor field parametrized as $\psi = \exp( i \theta) \left[ \exp( i \bar{\phi}/2)  , \, \exp( -i \bar{\phi}/2) \right]^T$, where for single particle states the external phase angle $\theta =  \theta_p \equiv  {\bf p } \cdot {\bf r}/\hbar $ is related to the momentum of the particle and the internal geometric phase $\bar{\phi} \equiv \phi =  \mathrm{tan}^{-1}(p_y/p_x)$ is identified with the polar angle and determines the direction of momentum flow in the plane. Note that $\bar{\phi}$ provides the correct half-angle property characteristic of a fermion field. Hence, $\bar{\phi}$ parametrizes a degenerate manifold for a given value of $|{\bf p}|$. In order to satisfy the condition $\langle \psi \rangle$ $=$ $0$, for a given energy eigenstate $\psi$ must form a superposition over all possible momentum directions, i.e., values of $\bar{\phi}$ must be randomly distributed wrapping around the polar angle $\phi$. The simplest way to satisfy this is for $\bar{\phi}$ to lie within the range $0 < \bar{\phi} <2 \pi (2 n +1 )$, where $n \in \mathbb{Z}^+$. The contribution from $n$ in the upper bound describes $2n$ bosonic vortices attached to the quark field with the Fermion geometric phase contained in the second term. Thus, the Fermi surface is in essence described by two orthogonal sets of momentum direction distributions. At each point in space these can be chosen to be the radially diverging (hedgehog) and rotational (vortex) flows, as depicted in Fig.~\ref{DiquarkBasis}(a). In each case the momentum angle wraps around the polar angle $2n$ times.

\subsection{Composite Quarks}
\label{CompQuarks}

\begin{figure}[]
\begin{center}
 \subfigure{
\label{fig:ex3-a}
\hspace{0in} \includegraphics[width=.5\textwidth]{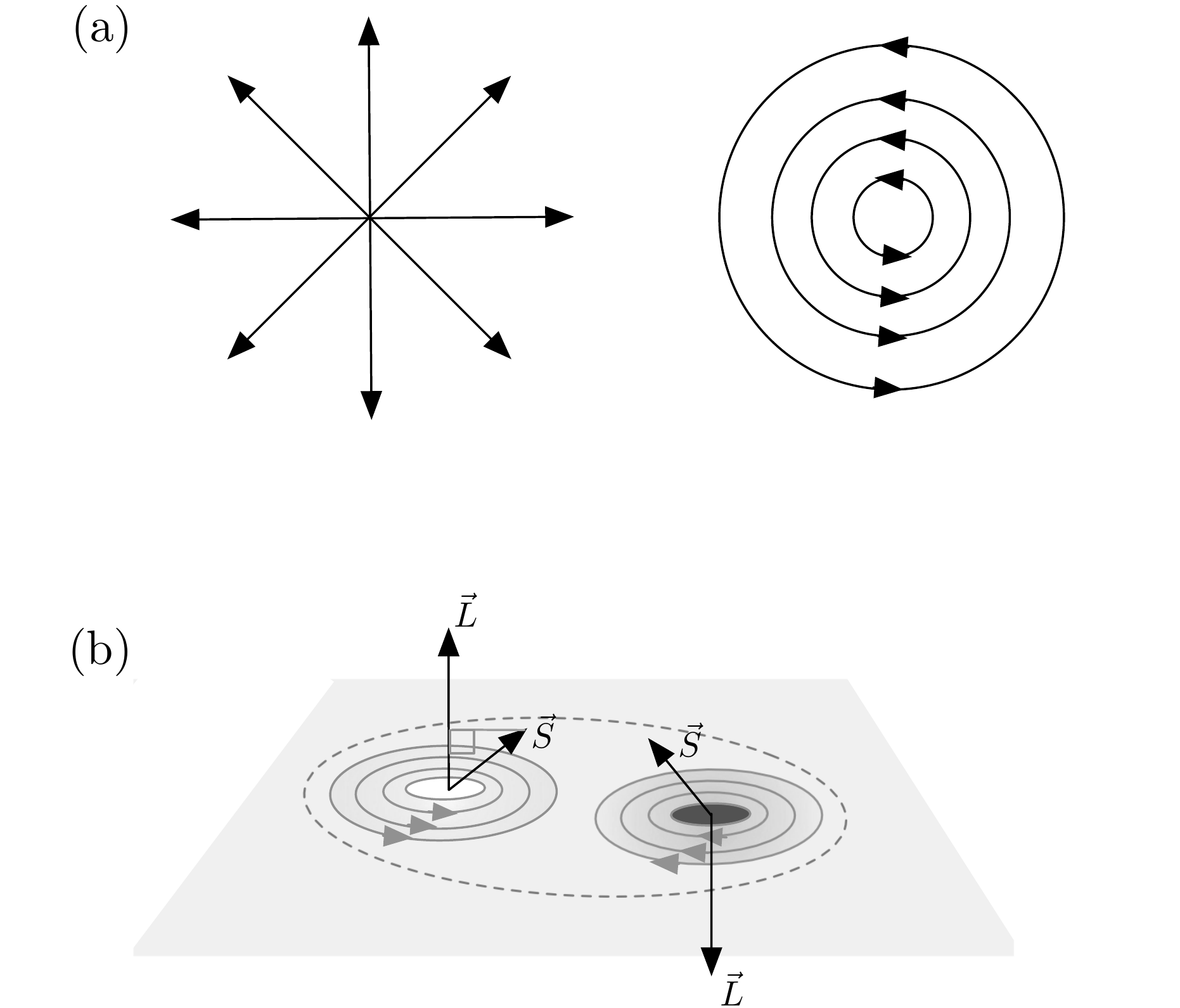}  } 
\vspace{0in}
\end{center}
\caption[]{(color online) \emph{Quasiparticle structure of the Gross-Neveu model with repulsive vector interaction.} (a) Local basis for single particle states comprised of superpositions of radial and vortical momentum states. The squeezed quark amplitude induces superpositions of different momentum directions at each point associated with a random internal phase of the spinor field. This effective introduction of topological order is required in order to maintain the fermion condition $\langle \psi \rangle = 0$. The extra phase winding describes an even number of bosonic vortices (orbital angular momentum $\vec{L}$) attached to a single quark (spin $\vec{S}$) creating a composite fermion, reminiscent of quantum Hall systems. (b) Density fluctuations over a uniform background expose a net extra phase winding, as adjacent fluid currents no longer cancel. Bound regions of surplus (light) or deficiency (dark) of fluid appear as oppositely rotating vortices. Vortex unbinding takes place when $m = m_c$, at which point $\vec{S}$ becomes polarized transverse to the plane, in the direction of $\vec{L}$. Note that the radial basis in (a) is associated with non-equilibrium time-dependent processes. }
\label{DiquarkBasis}
\end{figure}

Let us now see how quasiparticles emerge in this picture. Consider a density fluctuation over a uniform background in the vortical basis at the Fermi surface, depicted in Fig.~\ref{DiquarkBasis}(b). The resulting density gradient leaves some of the background vorticity ``exposed'' since contributions from neighboring fluid elements no longer cancel. As a result, a net current flows in the regions with elevated or depressed density, circulating in opposite directions in each case. This localized circulation reflects the statistics of the surrounding medium. That is, quasiparticles are fermions with bosonic vortices attached. This can be understood by recalling that the extra attached vortices were required in order to maintain overall Fermi statistics as the quark amplitude develops a finite expectation value, effectively forcing quarks into a densely packed squeezed state. Under such conditions the attached vortices provide bosonic screening between nearby fermions. The significance of this point can be better appreciated by recalling that fermions and bosons experience, respectively, repulsive and attractive exchange interactions. Hence, high density quarks are sparred the full impact of exchange repulsion by forming composite structures comprised of a bosonic vortex with a fermionic core, a more energetically favorable configuration than the bare fermion.$^{\footnotemark[1]}$\footnotetext[1]{The phenomenon of composite particles is quite familiar in quantum Hall systems. For example, see Ref.~\cite{Jain:2007} for a thorough treatment of composite fermions, and Ref.~\cite{Saarikoski2010} for an interesting overview of general composite particles in quantum droplets.}

\subsection{Characterization of Quasi-quarks}

In order to gain a more quantitative understanding of Fermi surface collective excitations, we proceed next to work in the hydrodynamic formalism through the canonical Madelung decomposition of the fermion field in terms of Weyl component amplitude and phase fields 
\begin{eqnarray} 
\hspace{-1pc} \psi( {\bf r}, t   ) = e^{i \vartheta( {\bf r}, t  )}   \eta(  {\bf r}, t  )   \left[  \mathrm{cos}\gamma( {\bf r}, t  )    , \,    e^{i \varphi( {\bf r}, t  )} \mathrm{sin}\gamma( {\bf r}, t  )  \right]^T \hspace{-.5pc} , \label{Madelung1}
\end{eqnarray}
where $\vartheta(  {\bf r}, t  )$ and $\varphi(  {\bf r}, t  )$ are the overall and relative phases, respectively, with the local spin-up (down) densities $\rho_{1(2)}(  {\bf r}, t  )$ related to the quark amplitude $\eta(  {\bf r}, t  )$ and chiral angle $\gamma(  {\bf r}, t  )$ by $\rho_1 = \eta^2  \mathrm{cos}^2\gamma$ and $\rho_2 =  \eta^2  \mathrm{sin}^2\gamma$. We will soon see that this choice of spinor decomposition is crucial in order to identify the underlying mechanism of the  QBKT transition. For convenience we now shift to using the total energy functional associated with our Lagrangian density. Incorporating the decomposition Eq.~(\ref{Madelung1}), then, after substantial algebraic manipulation, we extract the total energy for the Weyl spinor component fields
\begin{eqnarray}
 \hspace{-2pc} E \! &=& \! \! \!  \int \! d{\bf r} \left[ i  \, \eta^2  \, \mathrm{sin}2 \gamma \, {\bf n}_\varphi \cdot \nabla \vartheta   + \frac{1}{2}  \,  \mathrm{sin}2 \gamma  \, {\bf n}_\varphi \cdot \nabla \eta^2  \right. \label{landau1} \\
 && \left. \hspace{1pc} + \,   \eta^2   \mathrm{cos}2 \gamma  \, {\bf n}_\varphi \cdot \nabla \gamma +   i   \eta^2  \,\vert  {\bf n}_\varphi  \times  \nabla \gamma  \vert \right.  \nonumber  \\
      &&  \left.  \hspace{-.5pc} + \,   i  \frac{1}{2}  \,    \eta^2  \,  \mathrm{sin}2 \gamma \,   {\bf n}_\varphi \cdot \nabla \varphi  -  \frac{1}{2}   \,   \eta^2\,   \mathrm{sin}2 \gamma \,  \vert {\bf n}_\varphi  \times  \nabla \varphi \vert  \right. \nonumber \\
      && \left.   \hspace{2pc}+ \,   \eta^2  \,  {\bf n}_\gamma^T  {\bf M}  \, {\bf n}_\gamma   +   \frac{g^2}{2}  \,  \eta^4  \left( \mathrm{cos}^4\gamma  +    \mathrm{sin}^4\gamma \right) \right]  \, .   \nonumber 
\end{eqnarray}
Note that at very low temperatures, Eq.~(\ref{landau1}) is approximately equal to the Landau free energy (neglecting the entropy term). From here forward we will work in this time-independent formalism where covariance is not explicit but which favors a more physically intuitive perspective.

The gradient terms in Eq.~(\ref{landau1}) appear as scalar and vector products of the unit vector ${\bf n}_\varphi \equiv \left( \mathrm{cos}\varphi  \,,  \; \mathrm{sin}\varphi \right)$ with gradients of the various fields. These terms can be interpreted geometrically since the angle $\varphi$ describes the direction of the Dirac current. Hence, ${\bf n}_\varphi$ is a unit vector along that direction, which identifies the gradient terms in Eq.~(\ref{landau1}) as directional flows for the fields $\eta$ and $\gamma$ with respect to the quark momentum $\sim \nabla \vartheta$. The mass and interaction terms in the last line of Eq.~(\ref{landau1}) depend on both the overall density $\eta^2$ as well as the chiral field $\gamma$ through the spin polarization vector ${\bf n}_\gamma   \equiv \left( \mathrm{cos}\gamma  \,,  \; \mathrm{sin}\gamma  \right)$ and in the quartic interaction. The temperature dependent $2 \times 2$ mass matrix is defined as ${\bf M}(T) \equiv \mathrm{diag}\left[  m     - \tilde{\mu}(T) \, ,    - m     - \tilde{\mu}(T) \right]$, where we have indicated the chemical potential to be explicitly dependent on the temperature $T$.

To reveal the quantum BKT transition in our model, we focus on the broken chiral phase of the Fermi surface in Region III of Fig.~\ref{PseudospinDomain} ($T \ll T_c^{(2)}$) near the critical point $m \lesssim m_c$ and expand the massive fields near a minimum of the energy~Eq.~(\ref{landau1}). This yields the energy functional $E = E_0 + E_\mathrm{int}$ for the field fluctuations $\delta \eta$, $\delta \gamma$, $\vartheta$, and $\varphi$ given by
\begin{eqnarray}
      &&\hspace{-1pc}E_0 =  \int \! d{\bf r}  [   s_m   \delta \eta   \nabla_\parallel \delta \eta  +    i c_\eta s_m  \vert \nabla \vartheta  \vert \label{NematicLandau} \\
      &&\hspace{.5pc}+ \, c_\eta\left(  c_m  \nabla_\parallel \delta \gamma   +  i   \nabla_\perp \delta \gamma \right)   +   \frac{1}{2}  c_\eta  s_m  \left(     i           \nabla_\parallel  \varphi   -         \nabla_\perp \varphi  \right)   ]     \nonumber \, ,  \\
      &&\hspace{-1pc}E_\mathrm{int}=  \int \! d{\bf r}  \left[  8 |m_c|    \left(  \delta \eta^2  +  s_m  2 \vert m_c \vert  \delta \gamma^2  /g^2\right) \right.   \label{NematicLandauNext}   \\
      &&\hspace{.5pc}\left.- \, 4 s_m    |m - m_c| \,   \delta \gamma   \right]   \, ,  \nonumber   
\end{eqnarray}
where the coefficients are
\begin{eqnarray}
  \hspace{-1pc}  c_\eta &=& \left[  (2 \vert m_c \vert/g^2 )^{1/2} +   \delta \eta  \right]^2 \\
 \hspace{-1pc}c_m &=&  \left( 1 - 2 \delta \gamma^2 \right)  \frac{|m|}{|m_c|} \nonumber  \\
  &&-  \, 2 \delta \gamma  \,   \frac{1}{  |m_c|}  |m + m_c |^{1/2}  |m - m_c |^{1/2}       \, ,  \label{cm}\\
 \hspace{-1pc}s_m  &=&     \left( 1 - 2 \delta \gamma^2 \right)       \frac{1}{  |m_c|}  |m + m_c |^{1/2}  |m - m_c |^{1/2}   \nonumber   \\
 &&+ \,    2 \delta \gamma  \,  \frac{|m|}{|m_c|}  \label{sm}  \, .  
\end{eqnarray}
Expanding about a minimum of the energy shows the emergence of two massive fluctuations, $\delta \eta$ and $\delta \gamma$ in Eq.~(\ref{NematicLandauNext}), associated with radial and rotational directions in spin space, and two massless modes $\vartheta$ and $\varphi$. Classifying the fluctuations in our system, we find (parallel $\parallel$ and perpendicular $\perp$ notation referring to directions relative to the quark current): 1) a longitudinal spin density wave, or density compression wave, $\delta \eta$ parallel to the gradient of the quark phase $\vartheta$; 2) a massive spin-wave fluctuation in the local chirality encoded in the chiral fluctuation $\delta \gamma$; 3) a phase fluctuation, or spin wave, associated with  the quark current $\nabla \vartheta$; and 4) fluctuations $\varphi$ in the direction of the quark current. Encoded in $\nabla \varphi$ are the longitudinal ``snake'' mode $\nabla_\parallel \varphi$, which nucleates quantum turbulent flow in the quark bilinear condensates, and a transverse spin wave $\nabla_\perp \varphi$. This last mode introduces spatial divergence into the quark current, inducing local fluctuations in the quark density.$^{\footnotemark[2]}$ \footnotetext[2]{In the full dynamics associated with Eqs.~(\ref{NematicLandau})-(\ref{NematicLandauNext}) the parallel and traverse gradients in the phase $\varphi$ are coupled and describe time-dependent fluctuations in the quark momentum field $\nabla \vartheta$ akin to electromagnetic field propagation.}

The gradient terms in Eq.~(\ref{NematicLandau}) may be cast into a more enlightening form using geometric arguments. Specifically, fluctuations in the internal angle may be shown to convert to ${\bf n}_\varphi \cdot \nabla \varphi = |\nabla_\parallel \varphi | \to  |\nabla \times {\bf n}_\varphi|$ and $ \vert {\bf n}_\varphi  \times  \nabla \varphi \vert  = |\nabla_\perp \varphi | \to  |\nabla \cdot {\bf n}_\varphi|$; the former encapsulates orbital and spin currents; the latter accounts for time-dependent convection flow (see~\cite{bjorken64} for the Gordon decomposition of Dirac current and~\cite{Moses1959,Campolattaro1990} for the Weyl formulation of Maxwell's equations). To relate the $\varphi$ fluctuations to the divergence and curl of the direction field ${\bf n}_\varphi$ one expands the two orthogonal directional derivatives of $\varphi$ as follows 
\begin{eqnarray}
\nabla_\perp \varphi   &=&    \left( \mathcal{R}_{\pi/2}   {\bf n}_\varphi   \right)  \cdot \nabla \varphi   \\
 &=&   - \mathrm{sin} \varphi  \frac{\partial}{\partial x}  \varphi   +    \mathrm{cos} \varphi    \frac{\partial}{\partial y} \varphi    \\
   &=&  \nabla \cdot \left(  \mathrm{cos} \varphi  , \, \mathrm{sin} \varphi  \right)   \\
   &=&  \nabla \cdot {\bf n}_\varphi   \, , \label{divergence}
\end{eqnarray}
and 
\begin{eqnarray}
\nabla_\parallel \varphi     &=&    {\bf n}_\varphi   \cdot  \nabla \varphi  \\
 &=&   \mathrm{cos} \varphi  \frac{\partial}{\partial x}  \varphi   +    \mathrm{sin} \varphi    \frac{\partial}{\partial y} \varphi    \\
   &=&   \vert  \nabla \times \left(  \mathrm{cos} \varphi , \, \mathrm{sin} \varphi  \right)   \vert  \\
   &=&  \vert  \nabla \times  {\bf n}_\varphi  \vert \, . \label{curl}
\end{eqnarray}
Note that we have used the 2D rotation matrix $\mathcal{R}_{\pi/2}$ to rotate the direction field ${\bf n}_\varphi$ by $90$ degrees. Invoking the substitution $|\nabla_\perp   \varphi|    \;  \to \;  |\nabla \cdot {\bf n}_\varphi|$,  $|\nabla_\parallel    \varphi|   \;  \to \; \vert  \nabla \times {\bf n}_\varphi   \vert$ provides a more physically intuitive perspective. Indeed, as the field ${\bf n}_\varphi$ describes the direction of momentum flow, the divergence and curl forms here reveal precisely the same hedgehog and vortex structure presented in Sec.~\ref{CompQuarks} and illustrated in Fig.~\ref{DiquarkBasis}(a). We emphasize again that in the hydrodynamic picture, the curl of the direction field ${\bf n}_\varphi$ embodies the fermion spin and orbital angular momentum as well as additional quantized rotation of attached vortices, whereas the divergence of ${\bf n}_\varphi$ describes the convection part of the fermion current into or out of a volume element.

\section{Zero-temperature Berezinskii-Kosterlitz-Thouless phase transition}
\label{QBKT}

Since its inception, the BKT mechanism has found application in a wide range of physical systems characterized by a transition from algebraic to exponential behavior of long-range order~\cite{Berezinskii1971,Berezinskii1972,KT1973}. The topological nature of the BKT transition circumvents the well known Mermin-Wagner-Hohenberg no-go theorem forbidding finite-temperature spontaneous breaking of a continuous symmetry in dimensions $\le 2$~\cite{MermWag1966,Hoehenberg1967,Coleman1973}. Although in 2D spontaneous symmetry breaking at zero temperature is still possible. In the particular case of the holographic BKT transition for a D3/D5 brane system presented in~\cite{Jensen2010}, the underlying mechanism is rooted in the violation of the Breitenlohner-Freedman bound on the gravity side and the fact that parameters such as density and magnetic field have the same mass dimension.

In this section we derive the contribution to correlations in the superconducting phase of QCD coming from defects in the quark field, i.e., the composite quarks introduced in Sec.~\ref{CompQuarks}. As stated earlier, these defects incorporate the fundamental quark spin (Pauli vortex) and the extra attached bosonic vortices, both contained in the spin/orbital term of the Gordon decomposition of the Dirac current. By casting our problem in terms of a gas of such defects and integrating out the massive density and chiral modes, $\delta \eta$ and $\delta \gamma$, we will find that fluctuations in the former produce a hard-core delta function repulsion, whereas fluctuations in the latter provide a pairwise logarithmic attraction between defects whose coefficient vanishes at the critical point $m = m_c$. 

At ultra-low temperatures around a chosen ground state, to a good approximation, massive fluctuations can be treated as non-dynamical fields with each propagator contributing an overall factor of the inverse mass. Near a quantum phase transition, however, the chiral mode $\delta \gamma$ becomes massless decreasing as $m_\gamma \sim \sqrt{ m_c ( m_c -m)}$, in which case the corresponding propagator contributes a logarithmically divergent energy. Thus, at zero temperature the diquark dissociation energy at the critical point $m =m_c$ is supplied by long-wavelength quantum fluctuations in the chiral field. 

The general structure described here is precisely that of a 2D Coulomb gas with an additional infinite repulsion at the defect cores. Consequently, implementing the standard renormalization group analysis reveals the BKT signature exponential decay of quasi-long-range order for melting (when $T \ne 0$) or dissolving (for $T=0$) of the diquark condensate. The crucial point to emphasize here is that, in order to work, the precise mathematics that maps composite quarks to the Coulomb gas degrees of freedom requires a relativistic Dirac structure, as the latter is endowed with a peculiar chiral structure absent in ordinary BCS theory.

\subsection{Derivation of the QBKT Transition}

The effective energy for bound defects is obtained by Gaussian integration over the massive modes $\delta \eta$ and $\delta \gamma$ in two steps. First by integrating over the quadratic contribution from the $\delta \eta$ modes in Eqs.~(\ref{NematicLandau})-(\ref{NematicLandauNext}), then gathering the $\delta \gamma$ quadratic terms in the resulting expression and integrating over these. At each step the functional integral is performed by Fourier transforming to momentum space, then applying the standard Gaussian prescription (saddle-point approximation), re-exponentiating the resulting determinant, expanding about the extremum of the action, and finally re-expressing the result in real space. This procedure gives an effective theory for the spin wave (smooth) component of the quark current and a quark spin (defect) contribution from the vortical term. The defect contribution to the effective energy near criticality ($m \lesssim m_c$) is
\begin{eqnarray}
  \hspace{-.35pc}E    = \! \! \int \! \! d{\bf r}\,  d{\bf r}^\prime   \psi^*_d({\bf r}^\prime) \!   \left[  \rho_\eta   \delta({\bf r} - {\bf r}^\prime )   +  \rho_\gamma    \mathrm{ln}\left(  {\bf r} - {\bf r}^\prime    \right)                  \right]  \! \psi_d({\bf r})  \, , \, \label{DefectEnergy} 
  \end{eqnarray}
  with
  \begin{eqnarray}
  \hspace{-2pc} \hspace{1pc} \rho_\eta \equiv   \frac{  1}{64 m_c^3} (m + m_c ) (m_c -m)\,   , \;\;\; \rho_\gamma \equiv     \frac{m}{4 \pi m_c} \rho_\eta \, , \label{coefficients}
\end{eqnarray}
and where the defect field is given by $\psi_d = \hat{\bf \mathrm{z}} \cdot ( \nabla \times {\bf n}_\varphi)$. The logarithmic contribution in the second term of Eq.~(\ref{DefectEnergy}) is key, and comes from the long distance behavior in the momentum integration of the $\delta \gamma$ Green's function. Defect fields of the same chirality (opposite spin and momentum) are now coupled through a repulsive contact term with coefficient $\rho_\eta$, which comes from integrating out the amplitude field, and an attractive logarithmic term with coefficient $\rho_\gamma$ coming from the chiral field. Note that both coefficients in Eq.~(\ref{coefficients}) vanish at $m=m_c$ but at different rates. Diquark binding is depicted in Fig.~\ref{DiquarkTunneling}(a).

Contrasting this with the result far from criticality ($m \ll m_c$), we find that the second term in Eq.~(\ref{DefectEnergy}) is an attractive delta function in the separation $\delta(|{\bf r} - {\bf r}^\prime|)$, which reduces the strength of the hard-core repulsion coming from the first term. This happens since the mass of $\delta \gamma$ fluctuations in this regime is large compared to the characteristic low-temperature momentum. Thus, in addition to the unbinding that occurs for $m \ge m_c$, a cross-over occurs between two regimes: logarithmic binding around $m \lesssim m_c$ to unbound defects for $m \ll m_c$.

Equation~(\ref{DefectEnergy}) can be mapped to the 2D Coulomb gas as follows. Note first that for a closed path encircling delta function defect sources we have 
\begin{eqnarray}
&&\oint     {\bf n}_\varphi \cdot d {\bf s}  =   \int \!  \left( d{\bf r}  \, \hat{ \bf z} \right)  \cdot \nabla \times    {\bf n}_\varphi \\
 \Rightarrow    &&\nabla \times    {\bf n}_\varphi({\bf r})   =   2 \pi \, \hat{\bf z} \sum_i n_i^v \, \delta^2\! \left({\bf r} - {\bf r}_i  \right)\, , 
\end{eqnarray}
where $d {\bf s}$ is a tangent differential vector along the path. ``Magnetic'' charges for delta function defects at positions ${\bf r}_i$, ${\bf r}_j$, are denoted as $n_i^v$. Thus, in terms of an analog magnetic field notation $\nabla \times    {\bf n}_\varphi \equiv   \nabla \times {\bf \mathcal{B}}_d = \hat{ \bf z } \, \nabla^2 \mathcal{A}_d$, we then obtain $\nabla^2  \!  \mathcal{A}_d({\bf r})   =    \pi  \sum_i n_i^v \, \delta^2\! \left({\bf r} - {\bf r}_i  \right)$, which has the solution ${\bf \mathcal{A}}_d({\bf r})   =    \pi  \, \hat{\bf z}  \sum_i n_i^v \, \mathrm{ln} \left( {\bf r} - {\bf r}_i    \right)$. Incorporating these results into Eq.~(\ref{DefectEnergy}) and performing the integration over ${\bf r}$ and ${\bf r}^\prime$ gives 
\begin{eqnarray}
E=\pi^2  \sum_{i , j}   n_i^v n_j^v  \left[  \rho_\eta   \,  \delta^2 \! \left(   {\bf r}_i - {\bf r}_j    \right)      +      \rho_\gamma     \mathrm{ln} \! \left(  {\bf r}_i  - {\bf r}_j   \right) \right] \, , \label{CGE}
\end{eqnarray}
which leads to the diquark condensate correlation length obtained through standard renormalization techniques $\xi \sim  \exp \! \left[  c/ \sqrt{|m - m_c(T)|} \right]$, where $c$ is a constant and we have included finite temperature effects. The correlation length diverges exponentially near the critical point $m_c(T)$, consistent with the BKT theory: at $T=0$, a QBKT transition occurs at $m = m_c(0)$; for $T \ne 0$, $m_c(T)$ gives the critical point for a standard BKT transition, modified by quantum mechanical corrections.

 The zero-temperature transition displayed in Eq.~(\ref{CGE}) is driven by the quantity $\mu + A_0 - m$ (see Fig.~\ref{DiquarkTunneling}), which multiplies the logarithmic term and relates to the mass of chiral fluctuations $\delta \gamma$. The remarkable similarity of the second term in Eq.~(\ref{CGE}) to the free energy in conventional BKT for a system of size $R$ comprised of vortices with radius $a$ given by
 \begin{eqnarray}
 F = E - TS = (\kappa - 2 k_B T) \, \mathrm{ln}(R/a) \label{stndrdfree}
 \end{eqnarray}
 is evident by noting that the sum of chemical potential and vector density in our system measures the energy of an isolated defect, analogous to $\kappa$ in Eq.~(\ref{stndrdfree}), with the scalar mass analogous to the entropy factor $2 k_BT$. Hence, the ``thermal'' energy required to overcome vortex binding in the QBKT transition is evidently supplied by quantum fluctuations as $m$ approaches criticality. In Sec.~\ref{Chiral}, we will see that these fluctuations correspond to induced admixing of triplet pairs into the pure singlet Cooper pair states.

 \subsection{Analogy to 2D Magnetic Spin System}

\begin{figure}[]
\begin{center}
 \subfigure{
\label{fig:ex3-a}
\hspace{0in} \includegraphics[width=.375\textwidth]{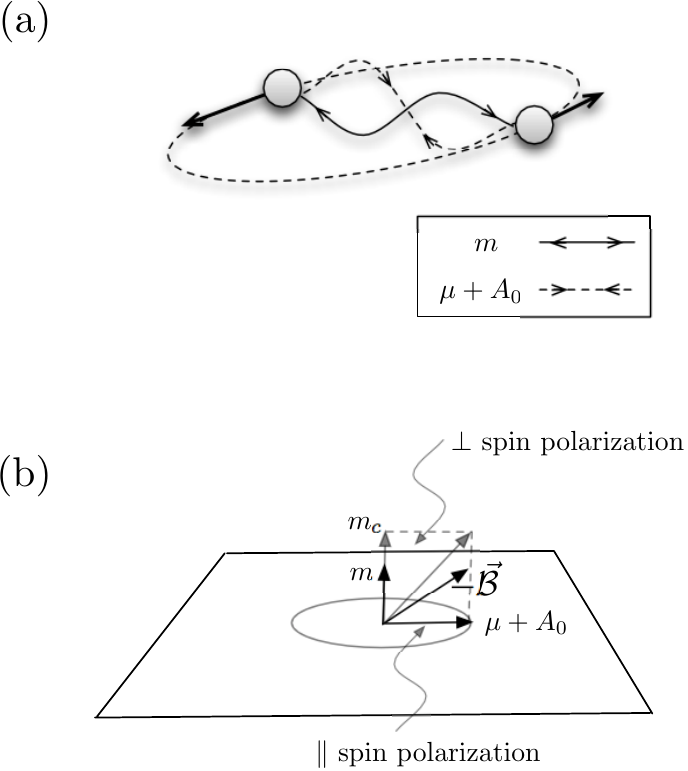}  } 
\vspace{0in}
\end{center}
\caption[]{(color online) \emph{Diquark in the Gross-Neveu model with repulsive vector interaction.} (a) The binding force between composite quarks arrises from the quark chemical potential $\mu$ and a mean field Hartree contribution from the time-like component of the vector interaction $A_0$. The scalar mass $m$ competes with these by adding repulsion between quarks, favoring chiral condensation instead. When $m = m_c = \mu + A_0$, unbinding of composite quarks is driven by a quantum Berezinskii-Kosterlitz-Thouless transition. (b) Binding of composite quarks cast in terms of the analog magnetic field $\vec{\mathcal{B}} \equiv  (\mathcal{B}_\parallel  , \, \mathcal{B}_\perp)^T$, where $\mathcal{B}_\parallel = -( \mu + A_0)$ and $\mathcal{B}_\perp =  - m$. For $|\mathcal{B}_\perp| < |\mathcal{B}_\parallel|$ the quark spin polarization is mixed between in-plane and transverse polarization. The QBKT transition occurs for $\mathcal{B}_\perp = \mathcal{B}_\parallel$. Note that the quark spin polarization is purely transverse for $|\mathcal{B}_\perp| > |\mathcal{B}_\parallel|$. In this sense, one can see that unbinding of composite quarks is analogous to conventional magnetic catalysis in systems with an applied external magnetic field.   }
\label{DiquarkTunneling}
\end{figure}

As we have shown, at zero temperature the Fermi surface can be classified according to the symmetry of the corresponding large $N$ semiclassical theory: formation of the Fermi surface breaks the chiral symmetry of the full finite temperature theory due to the sign change of the quadratic terms in Eq.~(\ref{brokensymmetry1}). Consequently, the quark spin acquires a well-defined polarization transverse to the plane analogous to the quantum spin problem in the presence of a transverse magnetic field at low temperatures.

In fact, the analogy is made evident by relating our parameters to a fictitious magnetic field defined as $\vec{\mathcal{B}} \equiv  (\mathcal{B}_\parallel  , \, \mathcal{B}_\perp)^T$, where $\mathcal{B}_\parallel = -( \mu + A_0)$ and $\mathcal{B}_\perp =  -m$, as shown in Fig.~\ref{DiquarkTunneling}(b). Under chiral symmetry breaking, the symmetry of the initial finite-temperature theory is inherited by the degeneracy of the ground state, i.e., the Fermi surface. In the case where the scalar mass $m=0$ ($\mathcal{B}_\perp =0$), the Fermi surface has a $\mathbb{Z}_4$ degeneracy, related to the number of minima in the effective potential for the quark field in Eq.~(\ref{brokensymmetry1}). This describes chiral symmetry of a Fermi surface comprised of two left-handed and two right-handed chiral ground states with the quark spin polarized along the plane. Tuning the mass away from zero, but well below the critical value $0 < m < m_c$ (analogous to rotating $\vec{\mathcal{B}}$ out of the plane), breaks the Fermi surface degeneracy to $\mathbb{Z}_2 \otimes \mathbb{Z}_2$. Here, what is initially a balanced superposition of transverse spin polarizations, equivalent to a well-defined in-plane chirality, is subsequently broken as the quark field is driven towards an admixture of left and right chiral states. When $m \ge  m_c = \mu + A_0$, a final round of chiral symmetry breaking occurs to a ground state with $\mathbb{Z}_2$ degeneracy associated with spin polarized in the positive $\hat{\bf z}$ direction, i.e., an equal superposition of left and right chirality for the quark field.

Thus, at $T=0$ tuning the ratio $m/(\mu + A_0)$ (or $\mathcal{B}_\perp/\mathcal{B}_\parallel$ in our analogy) through unity, dissolves the diquark condensate through a mechanism analogous to magnetic catalysis, which has been studied extensively for relativistic fermions in the presence of an external applied magnetic field (see for example Ref.~\cite{Miransky2015} and references therein).

\subsection{Chiral Fluctuations}
\label{Chiral}

The quantum phase transition at $m=m_c$ is driven by diverging quantum fluctuations in the chiral field $\delta \gamma$. To explain precisely what we mean by ``chiral field'' in the present context consider first that in (2+1)-dimensions the right and left massless chiral spinors can be expressed as $\psi_{R, L} = \rho^{1/2}   \exp(i \theta ) \left[ 1 , \, \pm \exp(i \phi ) \right]^T$, with the positive (negative) sign associated with the right (left) handed spinor, and $\phi \equiv \mathrm{tan}^{-1}(p_y/p_x)$. In 2D, the group SU(2) gets reduced to U(1) $\otimes$ Spin(2) with only two of the Pauli matrices $\sigma_{1, 2}$ playing a dynamical role. The equations for left and right hand spinors decouple with the third Pauli matrix $\sigma_3$ acting externally as a discrete symmetry transformation $\psi_L \leftrightarrow \psi_R$.

\begin{figure}[]
\begin{center}
 \subfigure{
\label{fig:ex3-a}
\hspace{0in} \includegraphics[width=.25\textwidth]{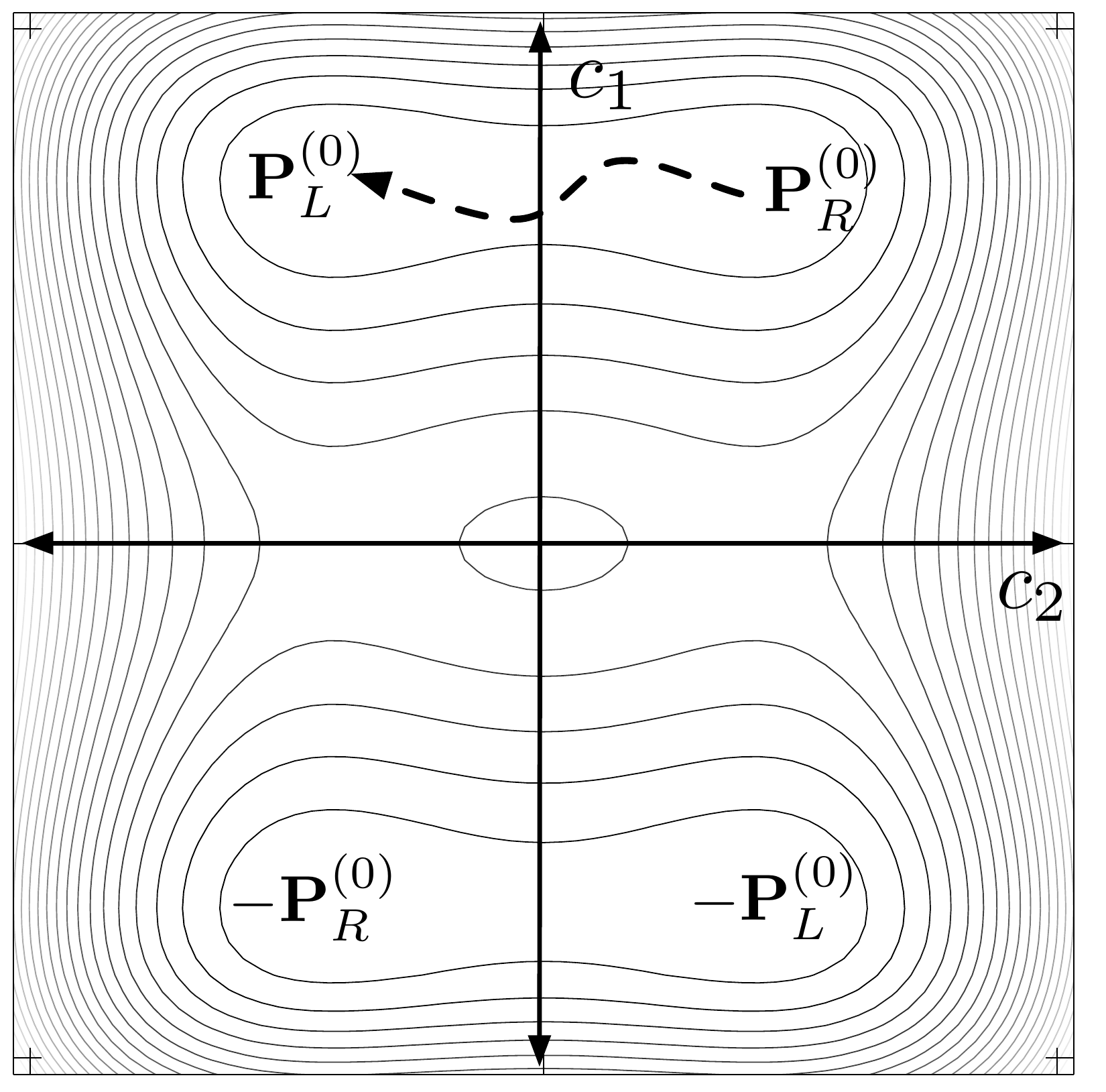}  } 
\vspace{0in}
\end{center}
\caption[]{(color online) \emph{Quantum tunneling between adjacent ground states of the effective potential. } Tunneling in spin space is depicted from the ${\bf P}^{(0)}_R$ to the ${\bf P}^{(0)}_L$ ground state driven by the $\delta \gamma$ field. As the quark mass is tuned $m \to m_c$ the two potential minima approach each other (see depiction of region III in Fig.~\ref{PseudospinDomain}(b)) significantly increasing the tunneling rate. Consequently, near criticality, Cooper pairs undergo extreme fluctuations between the singlet and triplet bound state until finally dissociating at $m = m_c$. }
\label{Tunneling}
\end{figure}

The Madelung decomposition of the Weyl spinor field takes the form $\psi_{R, L}$$=$$\rho^{1/2}$ $\exp(i \theta )$$\left[ \mathrm{cos}\gamma , \, \pm \exp(i \phi ) \, \mathrm{sin}\gamma \right]^T$, where the various parameters are functions of the coordinates. Here, as discussed, the additional field $\gamma$ is generally massive and therefore non-dynamical at zero temperature (except near a phase transition). Its expectation value is related to the particular ground state of the system through the average polarization vector ${\bf P}  \equiv (\mathrm{cos}\langle\gamma \rangle, \, \mathrm{sin}\langle\gamma \rangle)^T \propto ( c_1, \, c_2 )^T$ from Sec.~\ref{FermiSurface}, where proportionality accounts for a normalization constant. For illustration consider that we are in the ${\bf P}^{(0)}_R = ( 1 , \, 1)^T$ ground state of the manifold with $\mathbb{Z}_4$ symmetry, which corresponds to $\langle \gamma \rangle= \pi/4$ in the right hand chiral ground state. Consider then increasing the quark mass by a small amount so that the ground state symmetry is broken $\mathbb{Z}_4 \to \mathbb{Z}_2 \otimes \mathbb{Z}_2$. To expand about the new minimum we take $\langle \gamma \rangle \to  \pi/4 + \delta \gamma$, so that 
\begin{eqnarray}
{\bf P} &\approx& \left[ \mathrm{cos}(\pi/4 + \delta \gamma) , \,  \mathrm{sin}(\pi/4 + \delta \gamma)\right]^T \\
&\approx&  (1 , \, 1)^T + \delta \gamma  \, ( 1 , \, -1 )^T  \\
&=& {\bf P}^{(0)}_R + \delta \gamma  \, {\bf P}^{(0)}_L \, . \label{SpinFluctuations}
\end{eqnarray}
Hence a small fluctuation $\delta \gamma$ around a particular ground state, here ${\bf P}^{(0)}_R$, introduces a component of the opposite chiral ground state ${\bf P}^{(0)}_L$. Stated in words, the fluctuation $\delta \gamma$ describes quantum tunneling into an adjacent ground state as shown in Fig.~\ref{Tunneling}.

Functional integration over the massive fields couples quarks of the same chirality, i.e., states of opposite momentum and spin polarization, forming singlet Cooper pairs. Chiral fluctuations in the second term of Eq.~(\ref{SpinFluctuations}) affect the overall spin of these Cooper pairs. Consider the diquark state $\langle \psi^\dagger_R \psi_R \rangle$. The second term in Eq.~(\ref{SpinFluctuations}) induces the fluctuations 
\begin{eqnarray}
\hspace{-1.25pc}  \delta \langle \psi^\dagger_R \psi_R \rangle  =     \langle \psi^\dagger_R \psi_L \rangle  \delta \gamma   +    \langle \psi^\dagger_L \psi_R \rangle  \delta \gamma   +    \langle \psi^\dagger_L \psi_L \rangle  \delta \gamma^2  \, , 
\end{eqnarray}
 from which one sees that first-order corrections in the chiral field $\delta \gamma$ comprise spin-triplet states (unstable in ordinary BCS theory) whereas the second-order term describes tunneling into the opposite singlet state. Thus, the dominant effect of the $\delta \gamma$ field is to drive the individual chiral quarks that make up the diquark towards mixed chiral states. By tuning $m \to m_c$, fluctuations in $\delta \gamma$ become more pronounced as adjacent minima in the effective potential are brought closer together (see Fig.~\ref{Tunneling}), reducing the height of the barrier between them and mixing the opposite chiral state into the quark field. Finally, the quark field becomes an equal admixture of left and right chiral states at $m= m_c$. From this discussion, unbinding of composite quarks at criticality is then seen to result from the instability of the triplet bound state. 

\section{QCD phase diagram}
\label{PhaseDiagram}

In the final part of our analysis we translate our results into the more familiar form in terms of the quark chemical potential $\mu$, which can be related directly to the baryon chemical potential or the quark density. The region of the QCD phase diagram that emerges is the transition region between hadronic matter, defined by $\langle \bar{q} q   \rangle > \langle q q \rangle$, and quark matter, $\langle \bar{q} q \rangle < \langle q q \rangle$, which is sometimes referred to as the coexistence phase of QCD~\cite{Fukushima2011,Weise2012}. This is intimately related to the idea of hadron-quark continuity, i.e., a smooth transition from superfluid/superconducting hadronic matter to superconducting quark matter.

Another way to view this region is in terms of what has been dubbed ``quarkyonic'' matter~\cite{McLerran2007}, wherein the fermionic sector is defined by a mixed system comprised of free quarks as well as baryons. Unique features of this phase include the possibility of chiral symmetry restoration in the presence of confinement. We show this to be true for our model: chiral restoration occurs at low temperatures and for large chemical potential. Our model also exhibits the effect of a baryonic ``skin'' at the Fermi surface, another characteristic of the quarkyonic phase characterized by a layer of baryons of thickness $\sim \Lambda_\mathrm{QCD}$ measured from the Fermi surface that interpolates smoothly into the pure quark phase residing deep within the Fermi sea.

As we have seen, this baryonic picture of the Fermi surface is realized in our model through the composite quasi-quarks derived in Sec.~\ref{CompQuarks}, the lowest of which are naturally identified with the lightest baryon states. In extracting the QCD phase diagram from our model, we must keep in mind that a Fermi sea can only form when the chemical potential exceeds the baryon mass $M$. This reasoning is crucial for identifying a lower energy bound for the viability of our model. Hence, from our analysis of the Fermi surface we find that the tri-critical point for the onset of superconductivity occurs at $\mu = M$.

\begin{figure}[b]
\centering
\subfigure{
\label{fig:ex3-a}
\includegraphics[width=.53\textwidth]{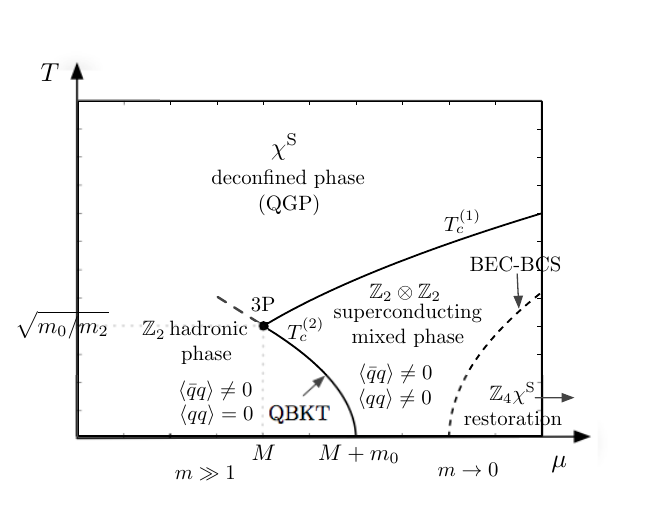}} \\
\caption[]{\emph{Temperature-chemical potential phase diagram for the Gross-Neveu model with repulsive vector term}. Chiral symmetry breaking phases are indicated based on Eqs.~(\ref{T1mu})-(\ref{T2mu}) associated with formation of the Fermi surface. The $\mathbb{Z}_2$-symmetric hadronic phase with vanishing diquark condensate lies to the left of the QBKT transition line. Just to the right of the QBKT line lies the superconducting phase described by a Coulomb gas of Cooper pairs with $\mathbb{Z}_2 \otimes \mathbb{Z}_2$ symmetry. For very large $\mu$ the system crosses over to free quarks indicated by the dashed curve to the lower right, obtained by shifting the lower curve in Fig.~\ref{FermiSurface} to smaller values of $m$. Interestingly, the BEC-BCS crossover occurs where color-flavor locking appears in more elaborate QCD models. The triple point (3P) occurs at the baryon mass $\mu = M$, marking the lower limit of viability for our model, with the dashed curve extrapolation for $\mu < M$ indicating the region where the associated inequalities (discussed in Sec.~\ref{FermiSurface} of the main text) cannot be satisfied. Note that $m_0$ is the constituent quark mass. Symmetries of the Fermi surface $\mathbb{Z}_2$, $\mathbb{Z}_2 \otimes \mathbb{Z}_2$, and $\mathbb{Z}_4$ are indicated to connect to the corresponding regions in Fig.~\ref{PseudospinDomain}: $\mu < M + m_0$ corresponds to the region $m > m_c$ in Fig.~\ref{PseudospinDomain} and $\mu > M + m_0$ to $m < m_c$. Directions of vanishing and increasing running quark mass $m$, $\mathbb{Z}_4$-chiral symmetry restoration, and deconfined chiral symmetric ($\chi^\mathrm{S}$) quark-gluon plasma (QGP) are shown. }
\label{BECBCS2}
\end{figure}

To construct the temperature-chemical potential phase diagram, we recast the boundary conditions from Sec.~\ref{FermiSurface} for the various chiral broken phases of the Fermi surface using a quadratic approximation for the scalar mass as a function of temperature $m(T) = m_0 - m_2 T^2$ and taking the chemical potential to be the free parameter. We then re-express the effective chemical potential in terms of the gauge potential $\tilde{\mu} = \mu - A_0$, and include the dynamical contribution to the gauge field through $A_0 = -(g^2/3)  \langle \psi^\dagger \psi \rangle + A_0^{(0)}$, with $A_0^{(0)} > 0$. This gives the chiral phase boundaries for the Fermi surface
\begin{eqnarray}
T^{(1)}_c(\mu) &=&\sqrt{(1/m_2) \left(  m_0 -  \mu + M \right) }  \, , \label{T1mu} \\
T^{(2)}_c(\mu) &=& \sqrt{(1/m_2) \left(  m_0 +  \mu - M \right) } \, , \label{T2mu}
\end{eqnarray}
where we have set the bare potential $A_0^{(0)} = M$ and taken $g^2  = 3  \epsilon_F/ \langle \psi^\dagger \psi \rangle$. The constraint on $A_0^{(0)}$ comes from the lower viable limit for our model: the Fermi surface cannot form below the baryon mass and the four inequalities in Sec.~\ref{FermiSurface} cannot be satisfied for $\mu$ below the triple point. The second condition equates $g^2/3$ with the prefactor of the density in $\epsilon_F$, a condition required in order to force the density dependence in Eqs.~(\ref{T1mu})-(\ref{T2mu}) to reside exclusively in $\mu$. Equations~(\ref{T1mu})-(\ref{T2mu}) give the boundary curves in the $T-\mu$ plane plotted in Fig.~\ref{BECBCS2}, where we have indicated the crossover region by a dashed curve towards the lower right corner of the plot. Note that the quark chemical potential is directly related to the quark density and baryon chemical potential by $\mu = \mu_B/3 \sim n_q$.

The resulting $T-\mu$ phase diagram at intermediate to high densities shown in Fig.~\ref{BECBCS2} is remarkably consistent with results obtained using more traditional approaches (see, for example, Fig.~1 and discussions in Ref.~\cite{Alford2008}). A key difference in our result is the appearance of a quantum BKT transition between the meson dominated ($\langle \bar{q} q \rangle \ne 0$) and the color superconducting ($\langle q q \rangle \ne 0$) regimes. At zero temperature this is driven by decreasing the quark chemical potential below a critical value equal to the baryon mass $M$ plus the constituent quark mass $\mu = M + m_0$. Finally, note that the triple point for the deconfined, hadronic, and superconducting phases occurs at $\mu = M$.

\section{Conclusion}
\label{Conclusion}

We have found that a quantum Berezinskii-Kosterlitz-Thouless phase transition separates the superconducting phase from the ordinary hadronic phase, within the Gross-Neveu model of planar QCD modified by an additional repulsive vector interaction. In particular, the transition is driven by reducing the quark chemical potential $\mu$ through a quantum critical point near the baryon mass $M$. At larger values of the chemical potential, the superconducting phase undergoes a BEC-BCS crossover in the region near the color-flavor locking transition that appears in more sophisticated models. The topological degrees of freedom associated with the QBKT unbinding mechanism are composite quarks near the Fermi surface, which we identify as excitations in the baryon skin that extends down from the Fermi surface. Significantly, our model reproduces the QCD phase diagram around the hadron-quark mixed phase, valid for $\mu > M$, obtained purely by analysis of Fermi surface modes, which we have characterized in extensive detail.

Several parallels are worth noting with regards to holographic BKT transitions, beyond the similarities in driving mechanisms expounded upon particularly in Sec.~\ref{QBKT} of the present work. First, we recall that at $T=0$ the chiral field $\gamma$ associated with the third Pauli matrix $\sigma_3$ in Eq.~(\ref{Madelung1}) is inactive for $\mu \gg \mu_c$, in which case the active modes are consistent with the number of spinor degrees of freedom in the 3D $\to$ 2D projection SU(2) $\to$ U(1) $\otimes$ Spin(2). When $\mu = \mu_c$, however, the vanishing mass $m_\gamma$ opens up a large time-like (in the 2D theory) length scale $\ell_\gamma \sim 1/p_\gamma$, recovering SU(2) symmetry through the now active generator $\sigma_3$. This picture at criticality is highly suggestive of the holographic-type BKT phase transitions~\cite{Jensen2010}, where the emerging third dimension in our problem is reminiscent of the radial direction in anti-de Sitter space that encodes the energy scale of the boundary field theory.

Another interesting point relates to Efimov states and their role in holographic BKT transitions. Such three-body bound states are expected to play a fundamental part in the holographic mechanism by way of an infinite tower of excitations near the two-body dissociation threshold~\cite{Iqbal2010}. It is curious that the Cooper pairs in our model are realized as Fermi surface bound states of baryon like modes, states resembling bosonic trimers that dissociate in the absence of one constituent boson. This similarity appears even more striking by the fact that our baryon like modes are in fact composite quarks for which our model predicts an infinite tower of bound states relating to the number of vortices attached to an elementary quark. More generally, in light of the conclusions drawn from our results in Sec.~\ref{QFluctuations} wherein compositeness emerges as an inevitable consequence of Dirac particles at high densities, we conjecture that a deeper relationship must exist between composite fermions in conventional condensed matter quantum Hall systems and the mathematical structure of Dirac spinors.

Confirming connections between holographic and four-fermion theories demands a rigorous construction of the dual gravitational description for our model. From a practical standpoint, though, this is somewhat academic as the primary appeal of holographic techniques is in circumventing the challenge of solving the boundary field theory directly. Indeed, the appeal of the present approach resides in the notion that a fortuitous choice of the correct degrees of freedom and associated interactions in the boundary theory provides the most insight into inherently non-perturbative problems.

\bibliography{QCD_Phases_Refs}

\end{document}